\documentclass[12pt]{article}

\usepackage{graphicx}% Include figure files
\begin{document}

\title{Polarization freezing of $10^4$ optically-cooled nuclear spins by coupling to a single electron\\}

\author
{E. A. Chekhovich$^{1,3}$, M. N. Makhonin$^{1}$, J.
Skiba-Szymanska$^{1,}$\footnote{Current address: Toshiba Research
Europe Limited, 260 Cambridge Science Park,Cambridge CB4 0WE, UK},\\
A. B. Krysa$^4$, V. D. Kulakovskii$^{3}$, V. I. Fal'ko$^{2}$,\\
M. S. Skolnick$^1$, A. I. Tartakovskii$^1$\\
\\
\normalsize{$^{1}$Department of Physics and Astronomy, University of Sheffield,}\\
\normalsize{Sheffield S3 7RH, UK}\\
\normalsize{$^{2}$Department of Physics, University of Lancaster, Lancaster LA1 4YB, UK}\\
\normalsize{$^{3}$Institute of Solid State Physics, Chernogolovka, 142432, Russia}\\
\normalsize{$^{4}$Department of Electronic and Electrical Engineering, University of Sheffield,}\\
\normalsize{Sheffield S1 3JD, UK} }
%\pacs{73.21.La, 75.75.+a, 78.55.Et}



\maketitle


\textbf{
The nature of the nano-scale environment presents a major challenge for solid-state implementation of spin-based qubits. In this work, a single electron spin in an optically pumped nanometer-sized III-V semiconductor quantum dot is used to control a macroscopic nuclear spin of several thousand nuclei, freezing its decay and leading to spin life-times exceeding 100 seconds at low temperatures. Few-millisecond-fast optical initialization of the nuclear spin is followed by a slow decay exhibiting random telegraph signals at long delay times, arising from low probability electron jumps out of the dot. The remarkably long spin life-time in a dot surrounded by a densely-packed nuclear spin environment arises from the Knight field created by the resident electron, which leads to suppression of nuclear spin depolarization.
}

%\%\textit{Introduction}

Spin-based approaches are one of the promising routes for quantum
information processing (QIP), for which an essential requirement
will be long-lived spin memory and coherence relying on effective
isolation of a spin-qubit from its magnetic environment. In this
respect, good candidates for solid-state implementation of QIP
are materials with well-isolated nuclear spins such as $^{29}$Si
nuclei \cite{DementyevSi,HayashiSi} and $^{31}$P impurities
\cite{P-in-Si,Kane} in silicon, nitrogen nuclei in N@C$_{60}$
molecules \cite{MorleyNC60} and $^{13}$C
\cite{Jelezko,Childress,Hanson} in diamond. In III-V
semiconductors, favored for fabrication of advanced quantum dot
(QDs) nano-structures suitable for both electrical \cite{Elzerman,electrostatic,electrostatic2} and optical \cite{Kroutvar,Atature,Greilich,Berezovsky} control of spin states, all atoms carry non-zero nuclear spin. This results in efficient spin decay via spin
non-conserving nuclear dipole-dipole interactions introducing an uncontrollable dephasing in QD-based qubits \cite{electrostatic,electrostatic2}. Dynamic nuclear polarization enabling control over the magnetic environment is a possible way
to circumvent this problem \cite{Imamoglu,Reilly,Gammon,Lai,Eble,Tartakovskii}.
Here we show that in a III-V semiconductor QD a single electron stabilizes the optically excited nuclear spin system leading to very long spin life-times of several hundred seconds, limited only by the electron residence time on the dot.

Our approach makes use of optical manipulation of a single
electron-charged InP self-assembled quantum dot embedded in bulk GaInP. Addressing individual well-isolated dots is crucial
for these studies to avoid sample inhomogeneities
\cite{Greilich,Oulton} and to operate with a precise knowledge of
the Overhauser field ($B_N$) on the dot. A typical dot
(4nm x 20nm x 20nm) contains about 10$^4$ nuclei and one trapped
electron originating from residual donors in the GaInP \cite{Skiba}. We
employ the technique of optically induced dynamic nuclear polarization and
photoluminescence (PL) measurements to monitor the temporal evolution of the Overhauser field at a temperature of 4.2K and magnetic field $B_z$ up
to 8T. We find nuclear polarizations which persist for hundreds of
seconds after the optical pumping has been turned off, with the
decay at long times having a random telegraph character,
which we attribute to random jumps of the trapped electron
from the dot. The trapped electron plays a crucial role: when the
electron occupies the dot, its inhomogeneous Knight field freezes
the nuclear spin decay mediated by the dipole-dipole interaction,
leading to the suppression of spreading of the nuclear
polarization from the dot into the surrounding
semiconductor.

The dynamic nuclear polarization occurs under circularly polarized
optical excitation of the dot leading, initially, to pumping of
the electron spin \cite{book}. The hyperfine
interaction leads to spin flip-flops between the electron and a single
nucleus\cite{book}. Macroscopic nuclear spin,
accumulated on the dot as a result of the dynamic nuclear
polarization, acts back on the confined electron with an effective
magnetic field $B_N$ \cite{book}. This results in the
modification of the electron energy spectrum manifested in the
splitting $\Delta E=\mu_B[g_hB_z-g_e(B_z + B_N)]$ observed in the electron-hole recombination spectrum [$g_{e(h)}$ is the electron (hole) g-factor].
The sign of $B_N$ depends on the helicity of the circularly polarized
excitation. In this work we measure $\Delta E$ to determine $B_N$,
directly reflecting the degree of the nuclear spin polarization on
the dot.

The probability for the spin transfer between the residual electron
and the nuclear spin system,
\begin{equation}\label{eq:ws}
w_s\propto|A_{hf}|^2/(\Delta E_e^2+\gamma^2/4),
\end{equation}
depends on the major energy cost of the electron-nuclear spin flip-flop event \cite{Erlingsson} - the electron Zeeman splitting $\Delta E_e=\mu_Bg_e(B_z+B_N)$.
$A_{hf}$ is the hyperfine interaction constant and $\gamma$
is the electron state broadening \cite{book}. The resonant form of the rate $w_s$ in Eq.\ref{eq:ws} results in feedback between the nuclear spin polarization on the dot and the spin transfer rate. In particular, an abrupt build-up of nuclear spin on the dot under optical excitation, the so-called  nuclear spin switch effect
\cite{Braun2,Tartakovskii,Skiba}, occurs when $\Delta
E_e\approx 0$ for  $B_N\approx -B_z$ leading to a sharp increase
of $w_s$.

Measurement of the nuclear spin life-time on the dot has been
carried out with a pump-probe method based on single-dot PL
detection. First, the dot is excited by the strong pump and then a PL
spectrum excited by a weak delayed probe is measured.  In
order to identify an optimum duration and optical power of the
probe, which does not perturb the system, the nuclear polarization
rise time is firstly measured in the experiment sketched in Fig.1A. First, a long $\sigma$-polarized
erasing pulse is used to destroy any nuclear polarization left from
the previous excitation cycle. This is followed by a pump
pulse with helicity opposite to that of the erase pulse. The
nuclear polarization generated by the pump is determined from the
spectral splitting $\Delta E$, measured during the last quarter of
the pump pulse. The erase/pump cycle is repeated several times to
improve the accuracy of the measured $\Delta E$ value. The polarization
build-up is then obtained by plotting the nuclear field $B_N$ as a function of the pump pulse duration $t_{pump}$.

The nuclear spin pumping dynamics strongly depend on the
polarization of the optical excitation. Fig.1B
shows the dependence of $B_N$ on the optical pumping time in a
charged QD at $B_z$=0.32~T, measured for both circular
polarizations of the pump. For $\sigma^{-}$-polarized pump a slow
change in $B_N$ is observed. In contrast, for the $\sigma^{+}$
pump $B_N$ abruptly reaches a constant value after
$t_{pump}$$\sim$0.2~s. This indicates the condition when $B_N$
compensates the external field $B_z$ and positive feedback of
the spin pumping on the nuclear spin transfer takes place
eventually leading to the nuclear spin switch
\cite{Braun2,Tartakovskii,Skiba}. In the $\sigma^+$ case the observed sharpness of the transition is limited by  averaging of the PL signal during the detection time $t_{det}=t_{pump}/4$. In
the case of $\sigma^{-}$-excitation negative feedback of the
nuclear polarization on the spin transfer rate plays a major role,
slowing down the spin pumping dynamics. In the limit of long $t_{pump}$ the
nuclear field reaches a steady-state value, $B_N^{cw}$, strongly
dependent on the polarization of incident light
due to the feedback described by Eq.\ref{eq:ws}. Such dependence
(Fig.1C) persists in a wide
range of magnetic fields $B_{z}<3$T, with $B_N^{cw}$ growing with $B_z$
from $B_N^{cw}$$\sim$0.2~T at $B_z$=0. It reaches its maximum
$\approx$1.2~T under $\sigma^{+}$~excitation at $B_z$$\approx$1.2~T,
when $B_z$ and $B_N$ compensate each other, thus leading to
enhanced electron-to-nuclei spin transfer rate $w_s$. In high
magnetic fields ($B_z>$3~T) the difference in $B_N^{cw}$ as well
as in the pumping dynamics for $\sigma^+$ and $\sigma^-$
excitation vanishes as the nuclear field becomes small compared to
the external field and does not lead to any significant feedback
on the spin transfer rate $w_s$ \cite{Skiba}. Due to a large
electron Zeeman splitting $\Delta E_e$ in this regime, the nuclear
spin pumping becomes less efficient and $B_N^{cw}$ decreases down
to $\approx$0.6~T for both polarizations of excitation.

Below we use the build-up time ($\tau_{build-up}$) required to reach
70$\%$ of the temporal dynamic range of  $B_N$ as a measure of the
nuclear spin rise time. The dependence of $\tau_{build-up}$ on
$B_z$ for $\sigma^+$ polarized pump is shown by the triangles in
Fig.2. The nuclear spin build-up time increases
from $\tau_{build-up}$$\approx$5~ms at $B_z$=0 to 2.5~s at
$B_z$=2~T. The very marked slowing down of the spin pumping dynamics
occurs due to the increase of the electron Zeeman splitting in the
external field $B_z$, as described above by Eq.~\ref{eq:ws}.
However, the saturation of $\tau_{build-up}$ at $B_z>$2~T observed
in Fig.2 is not explained by Eq.~\ref{eq:ws}. This
discrepancy might be evidence for several other processes
contributing to the dynamic nuclear polarization with strongly
differing magnetic field dependences.

Understanding of the build-up dynamics enables us to measure the nuclear spin decay using the pump-probe techniques shown schematically in the inset in
Fig.3A. The nuclear spin polarization was
induced by a circularly polarized pump pulse with duration
$t_{pump}\approx 10~\tau_{build-up}$, long enough to reach the
steady-state nuclear field $B_N^{cw}$. Photo-excitation was then
blocked by a mechanical shutter for a time $t_{delay}$. After the
delay, the nuclear polarization was measured in PL excited with a
short pulse (much shorter than $\tau_{build-up}$).

An example of the pump-probe curves measured in this way is shown
in Fig.3A. Here the dependences of $B_N$ on
$t_{delay}$ for both $\sigma^+$ and $\sigma^-$ polarized pump
pulses at $B_z$=8~T are shown. For $t_{delay}$$\le$100~s no
significant decay of the nuclear field
$|B_N^{pump}|$$\approx$0.8~T is
observed. For longer delays, the nuclear polarization fluctuates
from 0 to almost $B_N^{pump}$, significantly larger than the
experimental error $\sim$0.15~T. Such large fluctuations of $B_N$ with random telegraph character observed after long delays mean that the nuclear spin decay in
the QD is a random process and $B_N$ is not a continuous function
of $t_{delay}$. Mean-square fitting with a single exponential
functions, averaging over the abrupt jumps, leads to an estimate
for  $\tau_{dec}$ of $\approx$250~s (see solid lines in
Fig.3A). As shown in Fig.2 $\tau_{dec}$ (squares) is weakly dependent on $B_z$ and exceeds
100~s. At $B_z$=0 the longest delay time which can be measured in
our pump-probe set-up is $\tau_{delay}$$\approx$30~s (shown with a
horizontal bar in Fig.2), within which no
detectable decrease of the nuclear polarization on the dot was
found. We also have not observed any dependence on the pumping time of the nuclear depolarization time in the charged dots studied here, indicating the absence of nuclear spin diffusion.

We attribute the very long decay times we observe to
the role of the Knight shift induced by the residual electron in the charged dot. The
Knight field it creates on the dot can be measured directly from the magnetic
field dependence of the Overhauser splitting $\Delta E_N$=$\mu_B
g_e B_N$ at low $|B_z|<20$mT as shown in Fig.3B.
When an external magnetic field compensates the inhomogeneous
Knight field for a portion of the QD nuclear spins, the Zeeman
splitting becomes zero for these nuclei. This enhances spin
relaxation, observed as a partial decrease ($<$10\%) of the
Overhauser field. Excitation with the opposite circular
polarization creates a Knight field of the opposite sign. The
splitting $2 B_e$ between the minima of $\Delta E_N$ curves for
the two polarizations gives the Knight field $B_e\sim$3~mT. The
measured value of $B_e$ significantly exceeds the local field
$B_L$$\sim$0.1~mT \cite{book} created by the nuclear dipole-dipole
interaction. Thus we conclude that the electron spin generates an inhomogeneous
Knight field leading to a strong inhomogeneity in the nuclear
Zeeman splitting (see diagram in Fig.3C), which suppresses nuclear spin diffusion and relaxation due to the dipole-dipole interaction \cite{Deng}.

The abrupt step-like nature of the nuclear spin relaxation
with the form of a random telegraph signal is very different from that
observed in previously studied systems \cite{Maletinsky2,Makhonin,book,Paget}.  We observe large fluctuations $\Delta B_N$ comparable to the value of $B_N$ itself,
which suggests that the nuclear spin relaxation is triggered by a discrete
process, most probably electron hops out of the dot. In the absence of the electron, the stabilizing Knight field disappears enabling the nuclear spin relaxation due to the dipole-dipole interaction, leading in particular to nuclear spin diffusion  \cite{Makhonin}. Note, that a similar effect is
expected if an additional (second) electron is captured by the dot,
forming a spin singlet double-charge state. The proposed scenario explains the weak magnetic field dependence of $\tau_{dec}$, since the latter depends only on the average time for the dot recharging.

The spin decay times found in our sample are about $10^4$ times
longer than in Schottky-gated structures \cite{Maletinsky2}. This
occurs since the electron and nuclear spins in QDs in our sample
form a stable closed system, where the spin exchange with the
electron sea in the contact \cite{Maletinsky2} is eliminated.
The observed $\tau_{dec}$ up to 500~s is considerably longer than $\tau_{dec}$$\le$5~s in neutral InGaAs dots placed in high magnetic fields, where the nuclear spin diffusion was found to be the main mechanism of the nuclear spin relaxation\cite{Makhonin}. The lower limit of $\tau_{dec}\approx$30~s observed in our work for  $B$=0 is also more than one order of magnitude longer than in a neutral InGaAs dot measured at zero field in Ref.\cite{Maletinsky2}.

We also find that the Overhauser field produces a similar
stabilizing effect on the electron spin at zero external field. A
clear correlation between the electron spin-flip time and nuclear
spin polarization has been  observed. This arises from the
suppression of the electron spin-flip (caused by the hyperfine
interaction) for non-zero Overhauser field on the dot, as can be
expected from the strong dependence of the spin flip-flop rate
$w_s$ on the electron Zeeman splitting $\Delta E_e$ in
Eq.~\ref{eq:ws}.

%\%\textbf{conclusion}

Our results demonstrate previously un-revealed dynamic properties of
the strongly coupled spin-system of a single electron and an ensemble
of nuclei in a semiconductor nano-structure. In the charged quantum dots a single electron and $\approx 10^4$ nuclei form an almost closed system with a
conserved total spin. Optical control through the electron enables fast optical initialization and read-out, that combined with very
long spin life-times, leads to opportunities for the use of  this nano-system as
a resource for storage of spin-encoded information.

We thank R. Oulton and A. J. Ramsay for fruitful discussions. This
work was supported by the EPSRC grants EP/C54563X/1, EP/C545648/1,
Programme grants GR/S76076 and EP/G601642/1, and by the Royal
Society. \\
%\bibliography{bibl}

\textbf{\Large Methods\\}

{\it QD sample}: The sample containing InP/GaInP quantum dots was
grown on a GaAs substrate by metal organic vapor phase epitaxy
(MOVPE). In order to access single quantum dots we deposited an
opaque metal mask with 400~nm clear apertures. Although the sample
was not doped intentionally, the background doping is sufficient
for the majority of the dots to be negatively charged. The charge
state of the dots used in this study was verified in several ways
as explained in Ref. \cite{Skiba} in detail. The evidence of
negative charging is obtained from polarization-resolved
photoluminescence (PL) spectroscopy in external magnetic field
both perpendicular and parallel to the sample surface. At zero
magnetic field the PL spectrum of a negatively charged dot
consists of a single line. Magnetic field perpendicular to the
sample surface (Faraday geometry) splits this line into a
circularly polarized doublet. For magnetic field parallel to the
sample surface (Voigt geometry) two doublets with orthogonal
linear polarization are observed. The energy splitting between
these four lines is directly proportional to the magnitude of the
field. Such a spectral pattern in the Voigt geometry is
characteristic of dots containing a single charge, and is
different from that of neutral dots \cite{Skiba}. However, the
sign of the charge in the dot can not be determined in this way.
In order to prove the negative sign of the charging we use PL
spectroscopy under circularly polarized excitation. For close to
resonance excitation we observe negative circular polarization
(NCP): at high excitation rate PL of the dot is predominantly
circularly polarized with helicity opposite to that of the
exciting laser. This phenomenon is a fingerprint of negatively
doped dots and is attributed to electron spin memory
\cite{Oulton,Cortez}. Finally, the observed effect of the Knight
field gives direct evidence that the dot is charged with a single
electron.

{\it Experimental technique}: In all of the experiments the sample
was kept in a low-pressure He atmosphere and was mounted on a
metal plate in direct contact with liquid He at 4.2~K. Magnetic
field perpendicular to the samples surface up to 8~T was provided
by a superconducting magnet. In this work we studied quantum dots
emitting at $\approx$1.85~eV. We used semiconductor lasers
emitting below the GaInP band-gap at $\approx$1.9 eV for PL
excitation as well as for nuclear spin pumping. The exciting laser
was focused on the surface of the sample with an aspheric NA 0.65
lens, providing a focusing spot of $\approx$2~$\mu$m. The same
lens was used to collect PL, which was then dispersed by a 1~m
double monochromator coupled with a CCD camera. The accuracy of PL
peak spectral positions determined by the CCD pixel size was
$\approx$30~$\mu$eV. However, measurement of the center of mass
for each peak provided much better precision, and was used to
measure spectral splittings with accuracy of better than $1
\mu$eV. In pump-probe experiments mechanical shutters were used to
modulate both lasers and PL signals, providing time resolution of
$\approx$2~ms.

%\newpage

%\newpage

%\textbf{\Large Figure Captions\\}

\begin{figure}
\includegraphics[width=16cm]{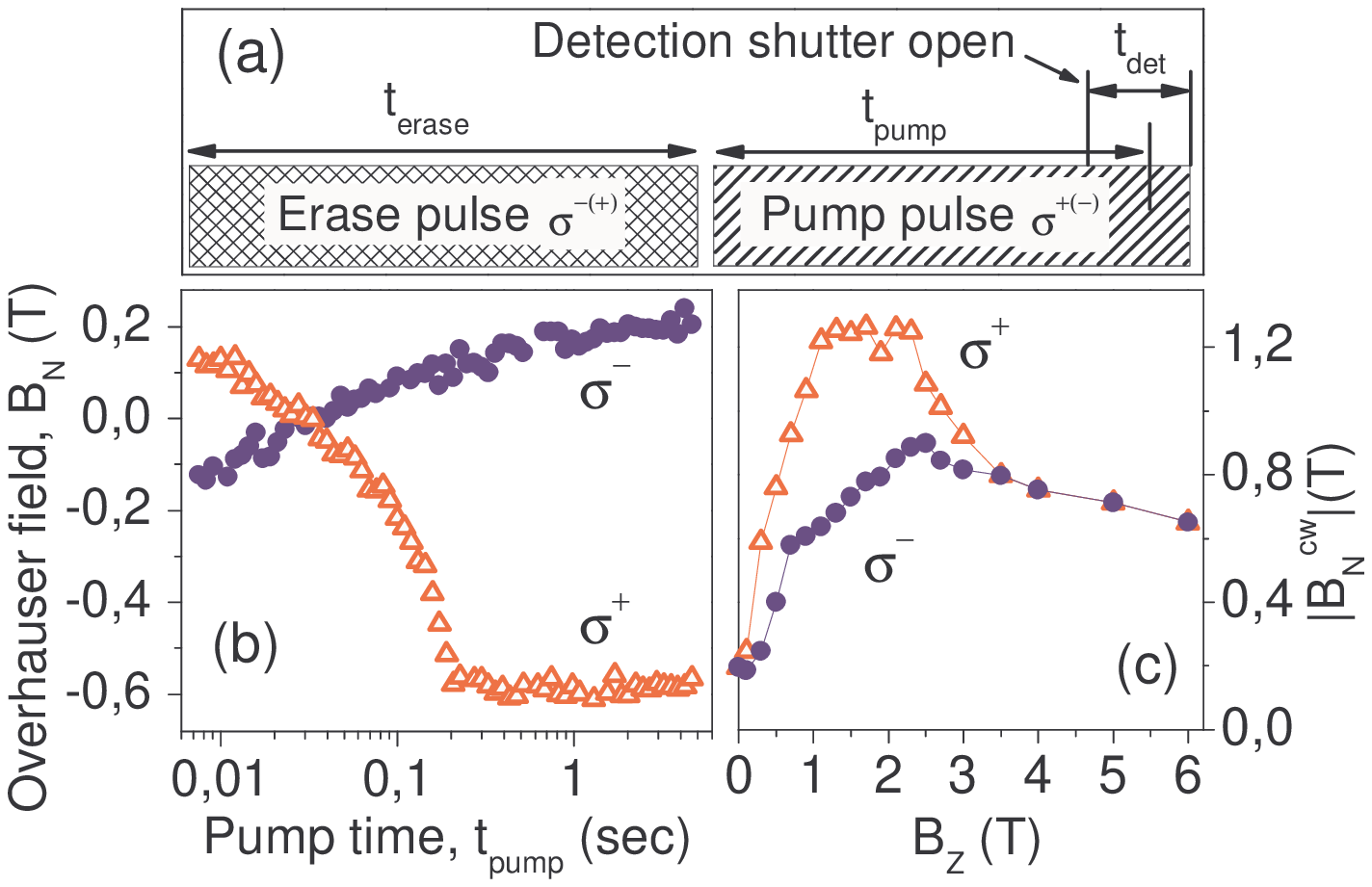}
\caption{Nuclear spin build-up dynamics in a
single negatively charged QD under circularly polarized optical
excitation. \textbf{(A)} Temporal diagram of a measurement cycle for
the nuclear spin build-up dynamics. A long circularly polarized
erase pulse erases the polarization left from the previous cycle.
Then a pump pulse of variable duration and of the opposite
helicity initializes nuclear polarization on the dot. The dynamics
of the Overhauser field build-up is obtained from the dependence
of the spectral splitting in dot PL, measured during $t_{det}$, on $t_{pump}$.
\textbf{(B)} Nuclear spin build-up dynamics under $\sigma^+$
(triangles) and $\sigma^-$ (circles) polarized excitation at
$B_z$=0.32~T. \textbf{(C)} Absolute magnitude of the steady-state
Overhauser field $|B_N|$ under $\sigma^+$ (triangles, $B_N$$<$0)
and $\sigma^-$ (circles, $B_N$$>$0) polarized excitation.}
\end{figure}

\begin{figure}
\includegraphics[width=16cm]{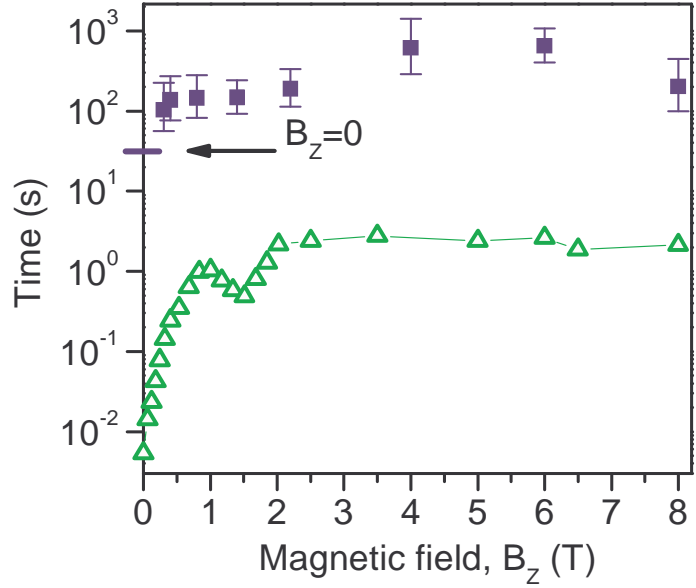}
\caption{Magnetic field dependence of the
optically induced nuclear spin dynamics. Triangles show the spin
build-up time under $\sigma^+$ polarized pumping. Squares show the
nuclear spin decay time once the pumping has been turned off. At
$B$=0 no decay of nuclear polarization is found for the longest
measured waiting time of 30~s (shown by the horizontal bar). Nuclear
spin build up is slowed down from $\tau_{build-up}$$\approx$5~ms
at $B$=0 to $\approx$2.5~s at $B$$>$3~T. At the same time the
nuclear spin decay time is almost independent of external field. }
\end{figure}

\begin{figure}
\includegraphics[width=16cm]{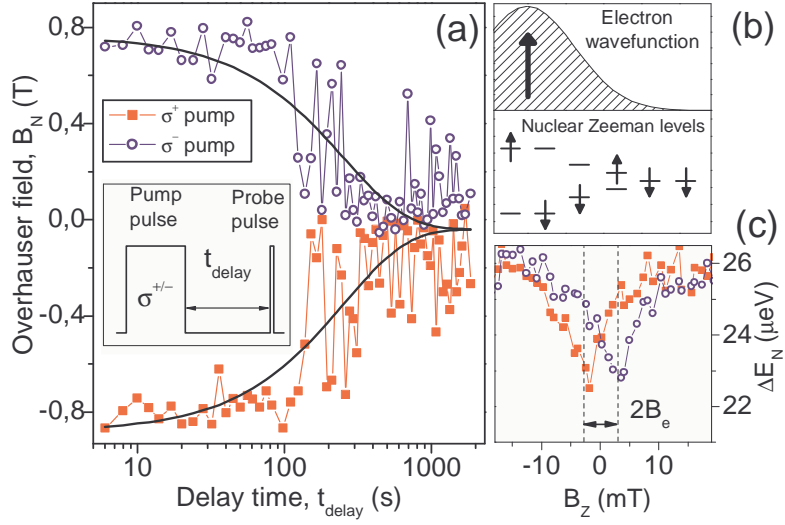}
\caption{\textbf{(A)}
Decay of nuclear spin polarization on the QD measured at $B_z$=8~T
in a pump-probe experiment schematically shown in the inset.
Squares and circles show data measured with $\sigma^+$ and
$\sigma^-$ polarized pumps respectively. For delays
$t_{delay}$$>$100~s large fluctuations of $B_N$ can be seen,
demonstrating that nuclear spin decay is a random discrete
process. The single exponential fit (shown with solid lines) gives an
estimate of the nuclear spin decay time
$\tau_{dec}$$\approx$250~s. \textbf{(B)} Schematic representation of the effect of a trapped electron on nuclear spin decay: inhomogeneous Knight
field causes energy splitting mismatch between different nuclei,
leading to suppression of nuclear spin diffusion out of the dot.
\textbf{(C)} Magnetic field dependence of spectral splitting
$\Delta E_N$ induced by nuclear polarization on the dot under
$\sigma^+$ (squares) and $\sigma^-$ (circles) excitation. The
observed minima of nuclear polarization correspond to a situation
when the external field $B_z$ compensates electron Knight field $\pm
B_e$ induced by $\sigma^{\pm}$ excitation, resulting in partial
depolarization of nuclei. The value of $|B_e|$$\approx$3~mT is
estimated from splitting between the minima.}
\end{figure}

\end{document}